\documentclass[12pt,halfline,a4paper]{ouparticle}
\usepackage{hyperref}
\hypersetup{hypertex=true,
    colorlinks=true,
    linkcolor=blue,
    anchorcolor=blue,
    citecolor=blue}

\begin{document}

\title{Distance to Default Based on the CEV-KMV Model}

\date{}

\author{Wen Su}

\abstract{This paper presents a new method to assess default risk based on applying the CEV process to the KMV model. We find that the volatility of the firm asset value may not be a constant, so we assume the firm's asset value dynamics are given by the CEV process $\frac{dV_A}{V_A} = \mu_A dt + \delta V_A^{\beta-1}dB$ and use the equivalent volatility method to estimate parameters. Focus on the distances to default, our CEV-KMV model fits the market better when forecasting the credit risk compared to the classical KMV model. Besides, The estimation results show the $\beta>1$ for non ST companies while $\beta<1$ for ST companies, which means their difference in the local volatility structure: ST volatility is decreasing with respect to the firm's asset while non ST volatility is increasing.}

\keywords{KMV Model, CEV Process, Distance to Default, Credit Risk}

\maketitle

\section{Introduction}
\label{sec1}

Due to the outbreak of the worldwide financial crisis in 2008, risk management especially credit risk management becomes one of the most centralized topics in finance. Bankrupt or default forecasting models, like the KMV model and the Credit Metrics model, now play a key role in assessing credit risk.

The well-known KMV model was developed by Kealhofer, McQuown, Vasicek and their company which is also called KMV Corporation in 1997, based on the seminal study of Merton (1974)\cite{Merton1974}. What KMV model contributes is it offers an insight that we can regard the equity as a call option on the firm’s asset. In the past several decades, the power of the KMV model in assessing credit risk has been tested. Kealhofer and Kurbat (2001)\cite{Kealhofer2001} focus on the listed companies in North America and find the KMV model captures all the information of the traditional accounting variables in financial statements when forecasting the default risk. Compared with Altman’s Z-Score (1968)\cite{Altman1968} and Ohlson’s O-Score (1980)\cite{Ohlson1980}, Hillegeist et al. (2004)\cite{Hillegeist2004} also argue that the KMV model provides significantly more information than either of the two accounting-based measures.

In order to improve accuracy of forecasting default probability, various advanced theories and methods are applied by other researchers. Bharath and Tyler (2008)\cite{Bharath2008} put up a naive predictor and find it performs slightly better in hazard models and in out-of-sample forecasts than both the KMV model and a reduced-form model that uses the same inputs. In order to find the best default point, Zhang and Shi (2016)\cite{Zhang2016} use particle swarm optimization and fuzzy clustering to modify the KMV model and the results show their FC-PSO-KMV model provides more accurate predictions.  Focusing on the default point, too, Song et al. (2020)\cite{Song2020} also use particle swarm optimization to adjust the KMV model since they find the classical KMV model could not do distinguish the ST companies.

The KMV model is established based on the Black-Scholes formula. Initially, Black and Scholes (1973)\cite{Black1973} assume the underlying stock returns dynamics are given by a GBM(Geometric Brownian Motion), i.e., constant volatility. However, the implied volatility smile, skew or smirk indicates that constant volatility may be wrong. More and more volatility models are developed in order to deal with volatility smile, like CEV(Constant Elasticity Variance) model proposed by Cox (1975)\cite{Cox1975}, Heston model (1993)\cite{Heston1993} and Hagan’s SABR(Stochastic Alpha Beta Rho) model (2002)\cite{Hagan2002}.

Recently, changing volatility seems to become a more and more welcomed consensus among researchers in financial economics, like Feng et al. (2018)\cite{Feng2018} find the smile of a company is more obvious than the Standard Poor Index and Li et al. (2021)\cite{Li2021} curves the shape of the smile of options on 50ETF in China. Besides, focus on CDS(Credit Default Swap) spread, Kelly et al. (2019)\cite{Kelly2019} define and construct a credit-implied volatility (CIV) surface.

It seems natural to apply non-constant volatility to KMV model, however, very few pay attention to modification on the GBM assumption in the KMV model, this paper will point out disadvantages of the GBM assumption and try to modify it.

The CEV model is considered to be an effective model which deals with the non-constant volatility. After Cox proposed the CEV model, plenty of researchers are taking further study and testing the power of the CEV model, like Beckers (1980)\cite{Beckers1980} and Davydov et al. (2001)\cite{Davydov2001}. Though we can see these non-constant volatility models including CEV model now play an important role in finance, yet still few study apply
the them to accessing credit risk. Our study will offer an insight for how to apply the non-constant volatility to the KMV model.

The remainder of this paper proceeds as follows. Section 2 introduces some preliminary methodology, including the KMV model, the CEV process and equivalent volatility. In Section 3, we describe how we collect the data and introduce some test approaches. In section 4 we first test the classical KMV model and analyse the volatility problem, then we propose a CEV-KMV model, and the empirical results show the CEV process does improve the forecasting ability of the KMV model. Finally, section 5 concludes our results and looks into the future.

\section{Methodology}
\label{sec2}

\subsection{The KMV Model}
Default forecasting is one of the most enduring themes of credit risk. In the well-known KMV model, the key point is computing default probability. 
Firstly, we assume that the firm's asset value ($V_A(t)$) dynamics are given by a GBM:
\begin{equation}
\frac{dV_A(t)}{V_A(t)} = \mu_A dt +  \sigma_A dB(t),\quad t\in [0,T],
\end{equation}
where $B(t)$ is a standard Brownian motion, $\mu_A$ and $\sigma_A$ represents the firm's asset value drift rate and volatility respectively. Let $D$ represents the debt of the company at a given maturity $T$, then naturally $V_A(T)<D$ means the company has no ability to afford its debt, which finally leads to default. In the KMV model $D$ is named as default point and usually chosen to be a half of the long-term debt (LTD) plus the short-term debt (STD) at $t=0$. The default probability can be expressed as
\begin{equation}\label{eqn:PD}
P\left( V_A\left( T \right) <D \right) =N\left( -\frac{\ln \frac{V_A\left( 0 \right)}{D}+\left( r-\frac{\sigma _{A}^{2}}{2} \right) T}{\sigma _A\sqrt{T}} \right) =N\left( -d_2 \right) ,
\end{equation}
where $N(\cdot)$ is the standard normal cumulative distribution function, $r$ is the risk free interest rate, SHIBOR(ShangHai Interbank Offered Rate) or $3\%$ may be a suitable choice in China market.

Unfortunately, both the firm's asset value $V_A(0)$ and the volatility $\sigma_A$ may be not easy to obtained, which makes it hard to compute equation (\ref{eqn:PD}). However, the volatility of the firm's equity can be estimated by using the historical stock data. In order to make use of this, we regard the firm's equity ($V_E$) as a call option on $V_A$ with strike price $D$, so $V_E$ can be expressed as, of course, Black-Scholes-Merton formula:
\begin{equation}
V_E(t) = V_A N(d_1) - e^{-r(T-t)}DN(d_2),
\end{equation}
where $d_2$ is similarly defined in equation (\ref{eqn:PD}) but $T$ should be taken place by $T-t$ and $d_1 = d_2 + \sigma_A \sqrt{T-t}$. Then $dV_E$ follows from It\^{o}’s lemma:
\begin{equation}
dV_E=\left( \frac{\partial V_E}{\partial t}+rV_A\frac{\partial V_E}{\partial V_A}+\frac{\sigma ^2V_{A}^{2}}{2}\frac{\partial ^2V_E}{\partial V_{A}^{2}} \right) dt+\sigma _AV_A\frac{\partial V_E}{\partial V_A}dB_t,
\end{equation}
implying that $V_E$ have local volatility
\begin{equation}
\sigma _E=\sigma _A\frac{V_A}{V_E}\frac{\partial V_E}{\partial V_A}=\sigma _A\frac{V_A}{V_E}N\left( d_1 \right) .
\end{equation}

Since $V_E(0)$ and $\sigma_E$ can be determined or estimated, $V_A(0)$ and $\sigma_A$ can be calculated by solving
\begin{equation}
\begin{cases}
	\sigma _E=\sigma _AN\left( d_1 \right) ,\\
	V_E\left( 0 \right) =V_AN\left( d_1 \right) -e^{-rT}DN\left( d_2 \right) ,\\
\end{cases}
\end{equation}
and then ($\sigma_A$,$V_A(0)$) is given by
\begin{equation}\label{eqn:KMVEQN}
\left( \hat{\sigma}_A,\hat{V}_A \right) =\underset{\left( \sigma _A,V_A \right)}{\mathrm{arg}\min}\left\{ \left( \sigma _E-\sigma _A\frac{V_A}{V_E}N\left( d_1 \right) \right) ^2+\left( 1-\frac{V_A}{V_E}N\left( d_1 \right) +e^{-rT}\frac{D}{V_E}N\left( d_2 \right) \right) ^2 \right\} .
\end{equation}

Besides, it is an interesting fact that rather than just pay attention to the theoretical default probability $N(-d_2)$, the KMV model defines a distance to default $\mathrm{DD}=d_2$ or
\begin{equation}
\mathrm{DD}'=\frac{V_A-D}{V_A\sigma _A}\approx \frac{\ln \frac{V_A}{D}+\left( r-\frac{\sigma _{A}^{2}}{2} \right)}{\sigma _A}=d_2,
\end{equation}
and then creates a map from DD to the historical expected default frequency (EDF). The more details about EDF can be found in Crosbie and Bohn (2019)\cite{Crosbie2019}. However, the rare default data in China do not allow us to do this, but it may bring some advantages in statistical inference for us to focus on the DDs.

\subsection{The CEV Model}
The CEV model was proposed by Cox (1975)\cite{Cox1975}. In this situation, we assume that the firm's asset value dynamics are given by 
\begin{equation}\label{eqn:CEVPro1}
\frac{dV_A\left( t \right)}{V_A\left( t \right)}=\mu_A dt+\delta V_{A}^{\beta-1}dB\left( t \right) ,\quad t\in \left[ 0,T \right] ,
\end{equation}
where $\delta>0$, $\beta>0$ are constant. The CEV process means $V_A$ has local volatility $\sigma_A = \delta V_A^{\beta-1}$.

From equation (\ref{eqn:CEVPro1}) we find if $\beta=1$ then CEV process turns back to the GBM. If $\beta>1$, the volatility is increasing with respect to the underlying. If $\beta<1$, the volatility is decreasing with respect to the underlying.

Similarly, if we know $\delta$ and $\beta$, it is theoretically available to compute default probability $P(V_A(T)<D)$. However, since $V_A(T)$ is not normal distributed, an analytical solution for default probability is hard to get, so the numerical solution may be a better choice.

By the Feynman-Kac Theorem, $g(t,x)=E[1_{\{V_A(T)<D\}}| \mathcal{F}_t]$ solves the partial differential equation (PDE):
\begin{equation}\label{eqn:CEVPDE}
\begin{cases}
	\frac{\partial u}{\partial t}+rx\frac{\partial u}{\partial x}+\frac{\delta ^2x^{2\beta}}{2}\frac{\partial ^2u}{\partial x^2}=0,&		t\in \left[ 0,T \right] ,\\
	u\left( T,x \right) =1_{\left\{ x <D \right\}},&		\\
\end{cases}
\end{equation}
it is very convenient for us to use the Finite Difference Method (FDM) to solve equation (\ref{eqn:CEVPDE}) and then
\begin{equation}
P(V_A(T)<D) = u(0,V_A(0)).
\end{equation}

\subsection{Equivalent Volatility}
Consider an European call option pricing problem, if the underlying asset value dynamics $dS$ (maybe not GBM) and terminal payoff $c(T,S)=(S_T-K)^+$ are given, consequently the option price $c(0,S)$ could be obtained theoretically. Define
\begin{equation}
d_1\left( \sigma \right) =\frac{\ln \frac{S}{K}+\left( r+\frac{\sigma ^2}{2} \right) T}{\sigma \sqrt{T}},\quad d_2\left( \sigma \right) =d_1\left( \sigma \right) -\sigma \sqrt{T}.
\end{equation}
Because of the 1-1 relation between Black-Scholes-Merton formula and implied volatility, option prices are often quoted by stating implied volatility. Thus the equivalent volatility $\sigma_B$ for $dS$ is defined as the unique solution of the following equation
\begin{equation}\label{eqn:EqVol}
c\left( 0,S \right) =SN\left( d_1\left( \sigma \right) \right) -e^{-rT}KN\left( d_2\left( \sigma \right) \right),
\end{equation}
where $c(0,S)$ is the option price calculated under assumption $dS$, and the “B” in the equivalent volatility $\sigma_B$ means the right hand side of equation (\ref{eqn:EqVol}) is Black's model (Black-Scholes-Merton model).

Hagan and Woodward (1999)\cite{Hagan1999} use singular perturbation methods to deduce the equivalent volatility formula for the CEV model:
\begin{equation}\label{eqn:CevVol}
\sigma _B=\frac{\delta}{f^{1-\beta}}\left\{ 1+\frac{\left( 1-\beta \right) \left( 2+\beta \right) \left( F-K \right) ^2}{24f^2}+\frac{\left( 1-\beta \right) ^2\delta ^2T}{24f^{2-2\beta}}+\cdots \right\},
\end{equation}
where $F=e^{rT}S_0$ is the future price and $f=\frac{1}{2}(F+K)$.

Obviously, if we assume the asset value dynamics $dV_A$ is given by the CEV process, then $\sigma_A$ computed by equation (\ref{eqn:KMVEQN}) is nothing but the equivalent volatility. So it is easy for us to estimate these parameters by methods like calibration, then the default probability can be computed by solving equation (\ref{eqn:CEVPDE}).

\section{Empirical Design}
\label{sec3}
In China, the listed companies under special treatment (ST) are always be considered as the ones with higher investment risk, while the non specially treated (non ST) listed companies usually have a lower investment risk. This type of equity risk is correlated with credit risk. Researchers have found a company with higher equity risk tends to have higher credit risk\cite{Chordia2017}\cite{Jiang2021}, like Friewald et al. (2014)\cite{Friewald2014} use the structural models show this positive correlation. So when testing a credit risk model, researchers in China always want to see if there exists a difference between ST and non ST companies.

Therefore, in order to test the performance of the KMV model, we will use the data in China stock market to see if the distances to default of the ST companies are significantly less than those of non ST companies.

\subsection{Data}

In order to establish KMV model, We obtain quarterly data of 3148 companies from Choice Database. The data includes: Equity Value $V_E$ (hunderd million CNY), Equity Volatility $\sigma_E$, Long-Term Debt (LTD, hunderd million CNY), Short-Term Debt (STD, hunderd million CNY). Here we use the 24 months historical stock return volatility offered by Choice Database as $\sigma_E$ and straightly compute default point by $D = \mathrm{STD} + 0.5 \mathrm{LTD}$.

The in-sample period is 2019Q1-2020Q4, and the out-of-sample period is 2021Q1-2021Q4. The criterion for data collection and cleaning is:
\begin{enumerate}
    \item We collect the data on 2022 March 18th, the number of listed companies in the Shanghai and Shenzhen main board on 2022 March 18th is 3148.
    \item Because the STD and LTD in financial statements can be collected only every quarter, we also only focus on the last trade day in every quarter when collecting $\sigma_E$ and $V_E$.
    \item If one company has no available $\sigma_E$ or $V_E$ data in a certain quarter, then we drop it in this quarter.
    \item $\sigma_E$ is the 24 months historical stock return volatility offered straightly by Choice. In order to avoid the error because of too few trading days, we drop a company in a certain quarter if the volatility offered by Choice is greater than 100\%.
    \item If a company has available $V_E$ and $\sigma_E$ data, but the STD and LTD is missing, then we fill them by the nearest neighbor. If the data cannot be filled with the nearest neighbor, then we drop it in this quarter.
\end{enumerate}

Table \ref{Table:Des_Stats} reports the basic information (mean) of our samples. ($V_E$: hundred million CNY, $D$: hundred million CNY)

\begin{table}[]
\centering
\renewcommand\arraystretch{1.5}
\caption{The Basic Information (Mean) of Samples} \label{Table:Des_Stats}
\begin{tabular}{ccccccccc}
\hline
\multirow{2}{*}{Quarter} & \multicolumn{4}{c}{ST companies}                                                                                & \multicolumn{4}{c}{non ST companies}                                                                               \\ \cline{2-9} 
                    & \multicolumn{1}{c}{Obs} & \multicolumn{1}{c}{$V_E$} & \multicolumn{1}{c}{$D$} & \multicolumn{1}{c}{$\sigma_E$} & \multicolumn{1}{c}{Obs} & \multicolumn{1}{c}{$V_E$} & \multicolumn{1}{c}{$D$} & \multicolumn{1}{c}{$\sigma_E$} \\ \hline
2019Q1              & 75                     & 39.824                   & 25.481                  & 0.480                   & 2367                   & 150.684                  & 125.747                 & 0.406                   \\
2019Q2              & 113                    & 27.696                   & 32.578                  & 0.506                   & 2379                   & 141.466                  & 126.212                 & 0.400                   \\
2019Q3              & 117                    & 29.414                   & 35.961                  & 0.499                   & 2440                   & 139.640                  & 123.388                 & 0.394                   \\
2019Q4              & 118                    & 33.043                   & 32.684                  & 0.506                   & 2473                   & 150.294                  & 131.179                 & 0.394                   \\
2020Q1              & 119                    & 31.668                   & 29.934                  & 0.512                   & 2494                   & 138.336                  & 122.793                 & 0.400                   \\
2020Q2              & 181                    & 27.968                   & 28.958                  & 0.504                   & 2430                   & 162.504                  & 129.883                 & 0.398                   \\
2020Q3              & 190                    & 33.341                   & 31.517                  & 0.504                   & 2460                   & 173.607                  & 130.027                 & 0.417                   \\
2020Q4              & 189                    & 39.659                   & 29.089                  & 0.493                   & 2487                   & 191.629                  & 130.857                 & 0.418                   \\
2021Q1              & 183                    & 46.499                   & 44.371                  & 0.457                   & 2479                   & 188.738                  & 124.845                 & 0.402                   \\
2021Q2              & 155                    & 34.073                   & 45.213                  & 0.486                   & 2523                   & 204.041                  & 129.957                 & 0.410                   \\
2021Q3              & 153                    & 38.117                   & 45.724                  & 0.519                   & 2585                   & 203.446                  & 138.774                 & 0.434                   \\
2021Q4              & 150                    & 47.025                   & 46.413                  & 0.521                   & 2642                   & 211.564                  & 140.365                 & 0.449                   \\ \hline
\end{tabular}
\end{table}

It can be seen that at the end of 2021, the average equity value of non ST companies was 21.16 billion yuan, far greater than 4.7 billion yuan of ST companies. However, for the default point, although the default point of non ST companies is 14.04 billion yuan, which is still far more than the default point of ST companies, 4.64 billion yuan, it can be seen that its difference is obviously not as large as the difference of equity value. For the default structure, the average equity value of ST companies is significantly closer to or even lower than the average default point, while the average equity value of non ST companies is greater than the average default point. For volatility, the stock price volatility of non ST companies is about 40\%, while that of ST companies is about 50\%, which also reflects that ST companies are more unstable and risky than non ST companies.

\subsection{Test Approaches}
\label{sec3.2}

The classical test method for KMV model in empirical analysis is to use the two sample $t$-test or/and Wilcoxon rank sum test to see whether the distances to default show a significant difference between ST and non ST companies. This paper will abandon the $t$-test, but still use Wilcoxon rank sum test and supplement a forecasting accuracy.

Usually, the two sample $t$-test relies on the GBM assumption. Because $\ln V_A$ is Normal distributed, so the distance to default
\begin{equation}
\mathrm{DD}=\frac{\ln \frac{V_A}{D}+\left( r-\frac{\sigma ^2}{2} \right) T}{\sigma \sqrt{T}}
\end{equation}
is Normal distributed, too. Then the two sample $t$-test can be the best method to see if the DDs of ST companies and non ST companies vary. However, the empirical result shows the DD is not Normal. It can be seen from Figure \ref{Fig:KMVDD} that Gamma or other skewed distributions can fit it far better than Normal.

In order to check the power of the KMV model, the Wilcoxon test may be helpful. By using only the information of the sample ranks, Wilcoxon test proposed by Wilcoxon is one of the most popular non parametric tests. Assume the ST DDs $X_1,X_2,\cdots,X_m$ from $F(x)$ and non ST DDs $Y_1,Y_2,\cdots,Y_n$ from $F(x+c)$. If we want to test
\begin{equation}\label{eqn:testnonpara}
H_0:c=0 \quad \mathrm{vs} \quad H_1: c>0,
\end{equation}
define $W_X$ is the sum ranks of the ST default distance $X$ in the hybrid sample, fixed the ratio $m/n$, when $n$ is large, under the null hypothesis the statistic
\begin{equation}\label{eqn:Z1}
Z_1=\frac{W_X-\frac{m\left( m+n+1 \right)}{2}}{\sqrt{\frac{mn\left( m+n+1 \right)}{12}}}
\end{equation}
is asymptotically standard normal. We reject the null hypothesis when $Z_1<c$, and it can be seen the less $Z_1$ is, the better the model assess the credit risk.

 Since the Pitman's asymptotic relative efficiency (ARE) of Wilcoxon test on a skewed distribution is far less than $1$, $Z_1$ is a really good way to test DD in this situation.

In addition, we define an accuracy statistics
\begin{equation}\label{eqn:Z2}
Z_2=\frac{\sum_{i=1}^m{\sum_{j=1}^n{1_{\left\{ X_i<Y_j \right\}}}}}{mn}.
\end{equation}
Because we are glad to see the ST DD is less than non ST DD, it's obvious that the model is better if $Z_2$ is closed to $1$. 

\section{Empirical Analysis}

\subsection{Empirical Results of The KMV Model}

Using equation (\ref{eqn:KMVEQN}) we can get $(\hat{\sigma}_A,\hat{V}_A)$, then the $\mathrm{DD} = d_2$ can be computed, too. Table \ref{Table:KMV_Result} summarizes the results of DD in KMV model and the corresponding $Z_1$, $Z_2$.

\begin{table}[]
\centering
\renewcommand\arraystretch{1.5}
\caption{KMV Model Results} \label{Table:KMV_Result}
\begin{tabular}{ccccccc}
\hline
\multirow{2}{*}{Quarter} & \multicolumn{2}{c}{ST} & \multicolumn{2}{c}{non ST} & \multirow{2}{*}{$Z_1$} & \multirow{2}{*}{$Z_2$} \\ \cline{2-5}
                    & Mean        & Std.       & Mean        & Std.        &                     &                     \\ \hline
2019Q1              & 5.001      & 2.763     & 5.275      & 2.272      & -1.502              & 0.562               \\
2019Q2              & 4.161      & 2.614     & 5.277      & 2.273      & -6.408              & 0.681               \\
2019Q3              & 4.086      & 2.487     & 5.356      & 2.286      & -7.148              & 0.698               \\
2019Q4              & 4.145      & 2.535     & 5.346      & 2.241      & -6.883              & 0.690               \\
2020Q1              & 4.035      & 2.357     & 5.268      & 2.291      & -6.733              & 0.685               \\
2020Q2              & 3.965      & 2.517     & 5.425      & 2.356      & -9.803              & 0.720               \\
2020Q3              & 4.090      & 2.580     & 5.154      & 2.240      & -8.337              & 0.683               \\
2020Q4              & 4.123      & 2.393     & 5.093      & 2.178      & -7.486              & 0.665               \\ \hline
\end{tabular}
\end{table}

According to the Table \ref{Table:KMV_Result}, except 2019Q1 we find the DD of ST companies is far less than that of non ST companies. The DD of ST companies is near 4 while that of non ST companies is greater than 5. We also find the standard deviations of ST companies DD are almost in $[2.5,2.7]$, while those of non ST companies DD are almost in $[2.1,2.3]$.

In the 8 quarters during 2019Q1-2020Q4, the most accurate prediction is offered by data in 2020Q2. In this quarter, the average DD of ST companies is $3.965$, while the average DD of non ST companies is $5.425$, and the corresponding $Z_1$ value of is $-9.803$ and $Z_2=0.720$, which means the accuracy of distinguishing ST companies from non ST companies by KMV model is 72.0\%. The worst prediction is offered by data in 2019Q1. In this quarter, the average DD of ST companies is $5.001$, while the average DD of non ST companies is $5.275$, and the corresponding $Z_1$ value of is $-1.502$ and $Z_2=0.562$, which means the accuracy of distinguishing ST companies from non ST companies by the KMV model is 56.2\%. For the other quarters, $Z_1$ is about $-7$, while the prediction accuracy is about 68.0\%.

\begin{figure}[]
\centering
\subfigure[ST companies]{
\includegraphics[width=3in]{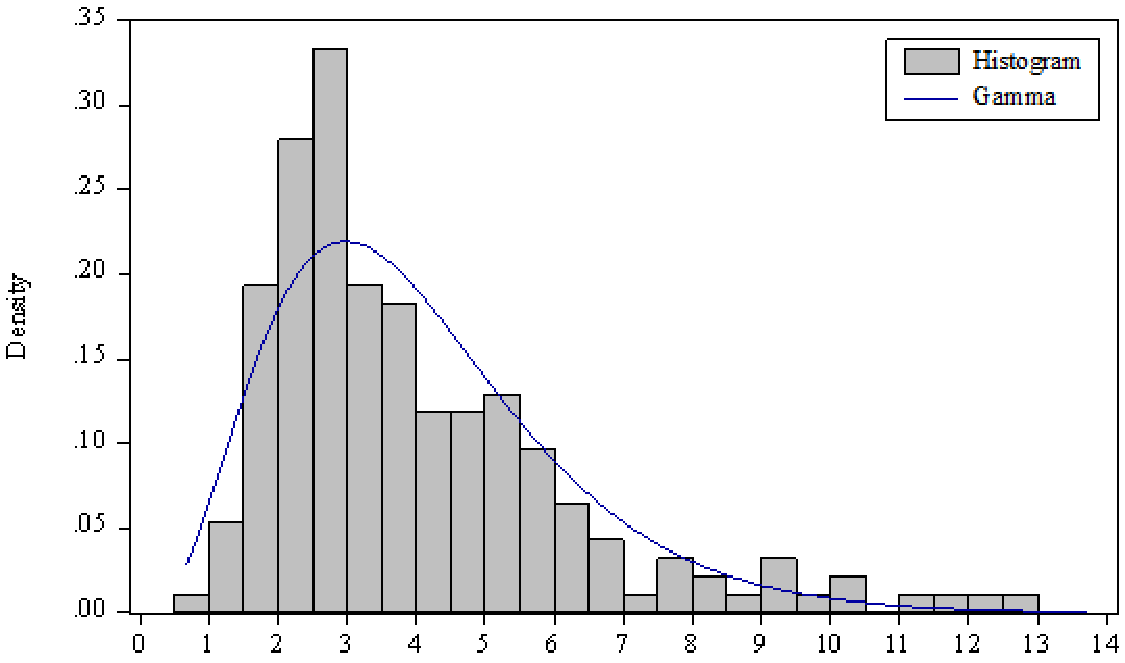}
}
\subfigure[non ST companies]{
\includegraphics[width=3in]{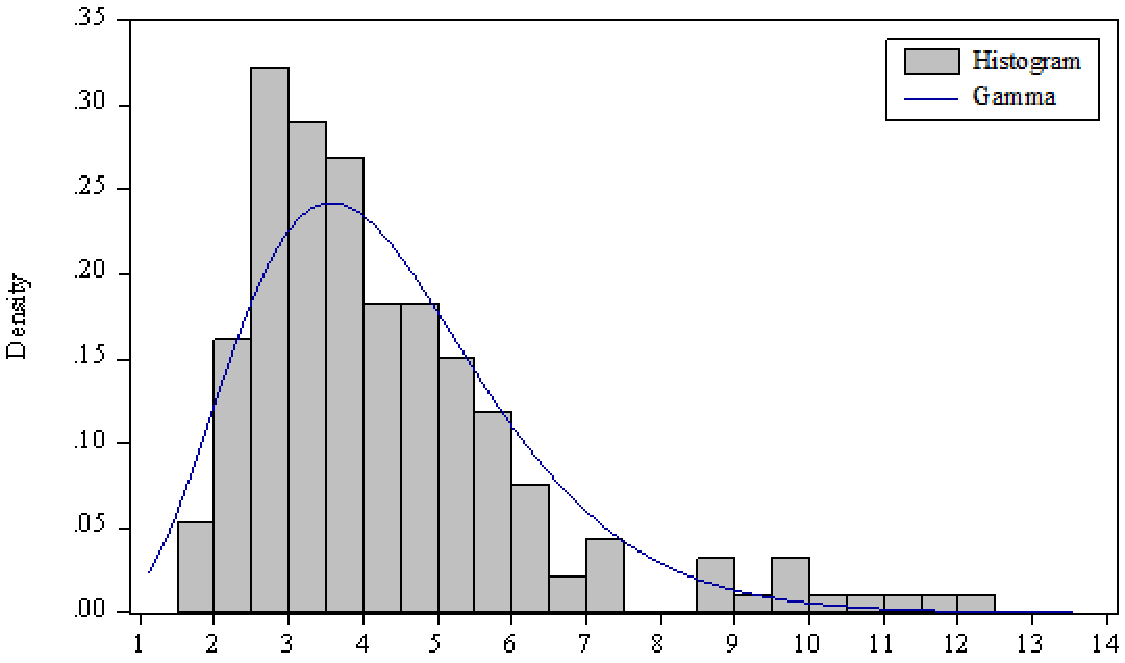}
}
\caption{Histogram and Gamma Density of DD (2020Q4)}
\label{Fig:KMVDD}
\end{figure}

Figure \ref{Fig:KMVDD} plots the histograms of ST DD and non ST DD computed by the KMV model, it's clear that Gamma density or other skewed unimodal distribution is better than the Normal. Besides, it seems that the non ST DD is more clustering near its mode, which is consistent with the larger standard deviation of ST DD shown in Table \ref{Table:KMV_Result}.

\subsection{Volatility Discussion}

Recently, some researchers have pointed out disadvantages of the KMV model, like Zhang and Shi (2018)\cite{Zhang2018} find that KMV model varies with firm size and what they pay attention to is the default
point. Actually, we focus on the relation between volatility and firm size: does volatility varies with the firm size?

\begin{figure}[]
\centering
\subfigure[ST companies]{
\includegraphics[width=6.2in]{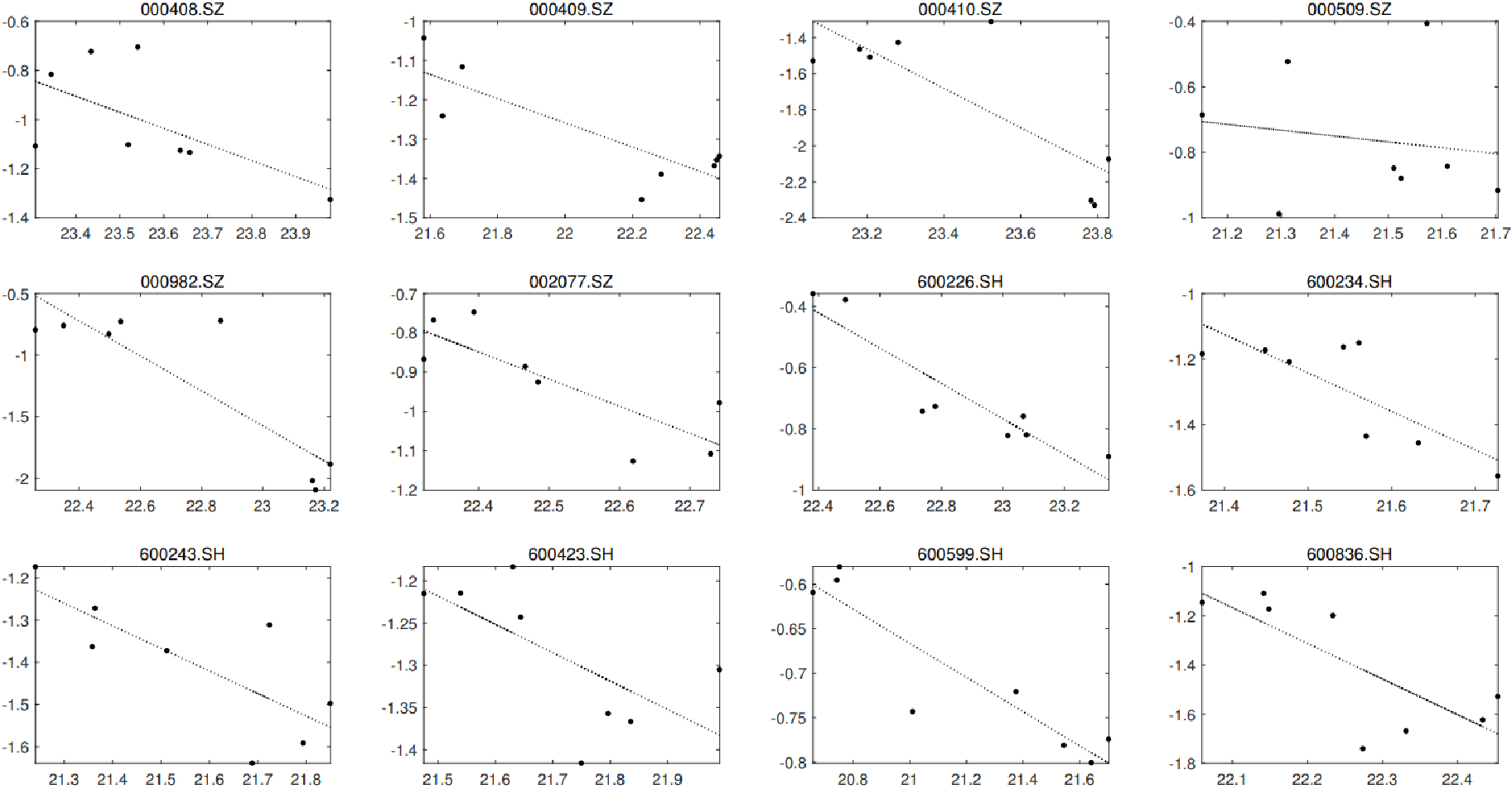}
}
\subfigure[non ST companies]{
\includegraphics[width=6.2in]{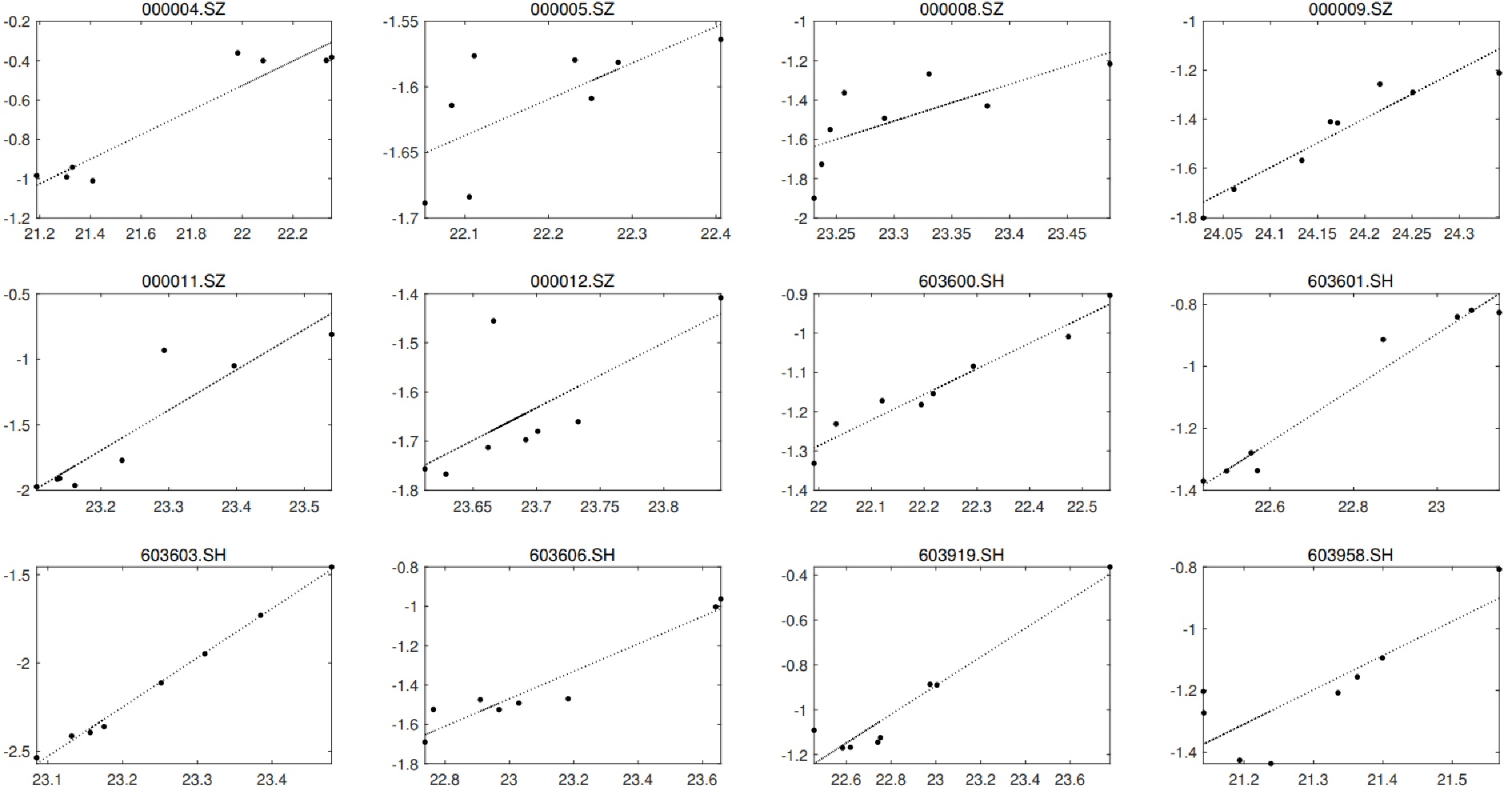}
}
\caption{The relation between $\ln V_A$ and $\ln \sigma_A$. The horizontal axis represents $\ln V_A$ and the vertical axis represents the $\ln \sigma_A$. The scatters are $(\ln V_A,\ln \sigma_A)$ given by the KMV model, and the dotted lines are fitted by linear regression.}
\label{Fig:SigVa}
\end{figure}

In theory, the volatility $\sigma_A$ in the KMV model is a constant. In practice, however, $\sigma_A$  may be not a constant. Figure \ref{Fig:SigVa} illustrates that at least with respect to $V_A$, volatility $\sigma_A$ is not a constant. Instead, $\sigma_A$ is negatively and positively correlated to $V_A$ in the picture (a) and (b), respectively.

Just as more and more financial economists have reached a consensus that “the constant volatility assumption is contrary to the implied volatility smile in the market”, it seems natural to consider non constant volatility in the KMV model. As shown in Figure \ref{Fig:SigVa}, if we assume
\begin{equation}\label{eqn:linear}
    \ln \sigma_A  = a + b \ln V_A,
\end{equation}
then we have
\begin{equation}\label{eqn:linear}
    \sigma_A  = e^a V_A^{b},
\end{equation}
which is exactly the local volatility under the CEV process. Hint from this, if the firm asset value dynamics are given by
\begin{equation}\label{eqn:CEVPro}
dV_A\left( t \right) =rV_A\left( t \right) dt+\delta V_{A}^{\beta}\left( t \right) dB_{t}^{Q},\quad t\in \left[ 0,T \right] ,
\end{equation}
then the prediction may be more accurate.

In the equation (\ref{eqn:CEVPro}), the local volatility of $V_A$ is
\begin{equation}
\sigma _A\left( V_A,t \right) =\delta V_{A}^{\beta -1}\left( t \right).
\end{equation}
If $\beta = 1$, then the CEV process turns back to the GBM. If $\beta>1$, then the local volatility is increasing with respect to $V_A$. If $\beta<1$, then the local volatility is decreasing with respect to $V_A$. From Figure \ref{Fig:SigVa}, it seems more ST companies have $\beta<1$ while more non ST companies have $\beta>1$.

Intuitively, using $a$ and $b$ from equation (\ref{eqn:linear}) then making $\delta = e^a$, $\beta = b + 1 $ is a fast way to estimate the parameters in equation (\ref{eqn:CEVPro}). But in fact, $\sigma_A$ is a constant volatility result obtained from the assumption of the KMV model, i.e. GBM, but is not the local volatility in equation (\ref{eqn:CEVPro}). Therefore, such estimation results have a model risk.

Actually, we can apply the idea of equivalent volatility, that is, for every company, assume $\sigma_A$ is the implied volatility in the market, and $\sigma_B(\delta, \beta)$ is the equivalent volatility under the CEV process. Under the square error loss, we can use $\sigma_A$ calculated from the data in the past 8 quarters to get the optimal $(\hat{\delta}, \hat{\beta})$, which eliminates the model risk in theory.

\begin{table}
\centering
\renewcommand\arraystretch{1.5}
\caption{$\beta$ Estimation in CEV Process} \label{Table:CEV_Est}
\begin{tabular}{ccccc}
\hline
    & 2021Q1 & 2021Q2 & 2021Q3 & 2021Q4 \\ \hline
non ST mean $\beta$  & 1.136  & 1.140  & 1.143  & 1.141  \\
ST mean $\beta$ & 0.981  & 0.977  & 0.973  & 0.971  \\ \hline
\end{tabular}
\end{table}

Table \ref{Table:CEV_Est} summarizes the average results of $\beta$ in the out-of-sample period estimated by rolling window and equivalent volatility. The estimated results seem to be very stable in the 4 quarters of 2021. For non ST companies, the average $\beta$ is $1.136$, $1.140$, $1.143$ and $1.141$, which are all greater than $1$, while for ST companies, the average $\beta$ is $0.981$, $0.977$, $0.973$ and $0.971$, which are all a little bit less than $1$, which is consistent with the previous empirical results.

\subsection{Empirical Results of The CEV-KMV Model}

Since this paper assumes that the asset value follows the CEV process based on the framework of the KMV model, so we call the hybrid model as the CEV-KMV model.

We have illustrated that the default probability under CEV process can be calculated by FDM (see equation (\ref{eqn:CEVPDE})), but in order to match the test approaches used previously and gain advantages in statistical inference, like we used to do in the classical KMV model, though $V_A(T)$ is not Normal distributed, we
can redefine the distance to default as
\begin{equation}
\mathrm{DD}_{\mathrm{CEV}} = N^{-1} \left(  P\left( V_A(T) < D\right)    \right).
\end{equation}
where $N^{-1}\left( \cdot \right)$ is the inverse function of the standard normal cumulative distribution function. Therefore, to test the performance of the CEV-KMV model, we need only estimate $\delta$ and $\beta$, which we have done in the last part. Table \ref{Table:CEV_Result} summarizes the average DD computed by the CEV-KMV model in the out-of-sample period (2021Q1-2021Q4), and compares them with the average DD of the KMV model. The corresponding $Z_1$ and $Z_2$ are also reported in the Table \ref{Table:CEV_Result}.

\begin{table}[]
\centering
\renewcommand\arraystretch{1.5}
\caption{Empirical Results of The CEV-KMV Model} \label{Table:CEV_Result}
\begin{tabular}{ccccccccc}
\hline
\multirow{2}{*}{Quarter} & \multicolumn{4}{c}{KMV}        & \multicolumn{4}{c}{CEV-KMV}    \\ \cline{2-9} 
                    & ST    & non ST   & $Z_1$     & $Z_2$    & ST    & non ST   & $Z_1$     & $Z_2$    \\ \hline
2021Q1              & 4.487 & 5.396 & -5.948 & 0.634 & 4.054 & 5.880 & -8.140 & 0.682 \\
2021Q2              & 4.176 & 5.363 & -7.356 & 0.678 & 3.936 & 5.986 & -8.869 & 0.714 \\
2021Q3              & 3.800 & 5.041 & -7.643 & 0.686 & 3.865 & 5.995 & -9.578 & 0.732 \\
2021Q4              & 3.889 & 4.966 & -6.595 & 0.662 & 4.014 & 6.120 & -9.075 & 0.722 \\ \hline
\end{tabular}
\end{table}

From the Table \ref{Table:CEV_Result}, it can be found that the ST DD calculated by our CEV-KMV model is basically similar to that calculated by the KMV model. In 2021Q1 and 2021Q2, the ST DD in the CEV-KMV model is lower, but in 2021Q3 and 2021Q4 the ST DD in the CEV-KMV model is higher. For non ST companies, the DD in the CEV-KMV model is significantly greater than that in the KMV model.
This is consistent with the estimation results in table \ref{Table:CEV_Est}. The average $\beta$ of ST companies is close to $1$, which means the CEV process of ST companies has little difference from the GBM, while the average $\beta$ of non ST companies is about $1.14$, far from the GBM. Besides, the statistics $Z_1$ and $Z_2$ manifest that the CEV process has improved the prediction ability of the KMV model. 

In the out-of-sample test, the Wilcoxon statistic $Z_1$ in the CEV-KMV model can reject the null at a higher significant level. The accurate statistic $Z_2$ increases to 68.2\%, 71.4\%, 73.2\% and 72.2\% from 63.4\%, 67.8\%, 68.6\% and 66.2\%.

\section{Conclusion}

Yet still no consensus exists in assessing default risk even if it has intrigued researchers for nearly 70 years. More and more tools and techniques have been used to assess default risk. This paper presents
some useful approaches to apply the CEV process to the KMV model. We focus on the distances to default and compare them under different models.

The first contributing factor we point out is that volatility of the firm asset value may be not a constant, and the figures suggest applying the CEV process instead of the GBM. We also notice that it is not easy to estimate the parameters in the CEV model, but we find the idea of equivalent volatility can help to achieve it. Finally, we find that both the Wilcoxon statistics $Z_1$ and the accuracy statiscs $Z_2$ manifest the CEV-KMV model can bring great advantages in distinguishing ST companies.

Besides, the fact we find that the monotonicity of local volatility vary with whether the company is ST is surprising. In general, for listed companies, since the greater volume and higher transaction frequency of a larger size firm must lead to a greater volatility (see in Weigand (1996)\cite{Weigand1996} and Lee et al. (2000)\cite{Lee2000}), most people may consider the larger firm size is, the greater the volatility becomes. However, we find many of ST companies do not follow this rule, the volatility of a ST company decreases when the firm size increases. We guess one of the main reasons for ST companies disobey the general rule may be the delisted
risk for ST companies. The less firm size is, the higher delisted probability will be. Extra risk often manifests a great volatility in stock market. In a word, this interesting fact we found may be caused by many other reasons and deserves further study.

\section*{Acknowledgements}
The author reports no conflicts of interest. The author alone is responsible for the content and writing of the paper.

\end{document}